\documentclass[aps,twocolumn,showpacs,superscriptaddress]{revtex4-1}
\usepackage{amsmath,amssymb}
\usepackage{graphics,graphicx}
\usepackage{dcolumn,bm}
\usepackage{psfrag}
\newcommand{\cu}
{\affiliation{Department of Physics, University of Calcutta,
92 Acharya Prafulla Chandra Road, Kolkata 700009, India.}}
\newcommand{\tifr}
{\affiliation{
Department of Theoretical Physics, Tata Institute of Fundamental Research, Homi Bhabha Road, Mumbai 400 005, India.}}

\begin{document}

\title{Exit probability in inflow dynamics: nonuniversality induced by range, asymmetry and fluctuation}

\author{Parna Roy}%
\cu
\author{Soham Biswas}%
\tifr
\author{Parongama Sen}%
\cu

\begin{abstract}

Probing deeper into the existing issues regarding the exit probability (EP)
in one dimensional dynamical models,  
we consider several  models where the 
states are represented by Ising spins  and the  information flows inwards. 
At zero temperature, these systems evolve to 
either of two absorbing states. The exit probability $E(x)$, which is the 
probability that the system ends up with all spins up starting with $x$ fraction of up spins 
 is found to have the general form  $E(x) = x^\alpha/\left[x^\alpha + (1-x)^\alpha\right]$.   The exit probability exponent $\alpha$ strongly depends on $r$, 
the range of interaction, the symmetry of the model and the induced  fluctuation.
Even in a nearest neighbour model, nonlinear form of EP can be obtained by controlling the fluctuations and for the same range, different
models give different results for $\alpha$.
Non-universal behaviour of the  exit probability is thus clearly established
 and the results are compared to
existing studies in models with outflow dynamics to distinguish the two
dynamical scenarios.

\end{abstract}


\maketitle

There are many systems in  condensed matter physics, magnetism, biology  
and social phenomena  \cite{bio,gen-dyn,book,soc2} which are found to reach an ordered 
state following certain dynamical rules. 
The dynamical rules represent the 
mechanisms by which macroscopic structures are generated from the microscopic
interactions.
The role of the dynamics is reflected in the scaling behaviour of relevant variables. Often we note power law scaling behaviour, e.g. in coarsening phenomena,
domains grow in a power law manner with time.
If two different dynamical schemes lead to  identical  
  behaviour of the 
relevant variables one may 
conclude that the two schemes are  actually 
equivalent. However,   careful studies are required to establish such equivalence.

Of late, a debate on whether inflow dynamics is different from outflow dynamics has emerged \cite{glauber,sznajd,krupa,cas}. Precisely, in  models involving spins, when the state of the central spin 
is dictated by its neighbours, it is a case of inflow of information. 
Outflow of information occurs when a group of neighbouring spins
 dictates  the state of all other spins neighbouring them.
To settle the debate, the exit probability (EP) is one of the features which is 
studied when the spins can be in up or down states. 
Starting with $x$ fraction of spins in the up  state, 
exit  probability $E(x)$ is the probability to reach a final state 
with all  
spins up. 

The Ising-Glauber model \cite{glauber} is an example of inflow dynamics where the local field
determines whether a spin will flip or not. 
An example where outflow of information takes place is  the Sznajd model \cite{sznajd}. 
In the  Ising-Glauber model,  a spin is selected 
randomly and
its state is updated following an energy minimisation scheme.
In one dimension, this always leads to either of two absorbing states: all spins up or all down.   
In the Sznajd model,
a plaquette of neighbouring spins is considered, if they agree
then the spins on the boundary of the plaquette are oriented along them. In one-dimension, the plaquette is  a panel of two spins. 
The Sznajd model has the same two absorbing states as in the Ising model. 
The two models also  have identical
exponents associated with domain growth and persistence behaviour during coarsening \cite{behera,stauffer1}.
However, a few other dynamic quantities were shown to be different for generalised  models with inflow and outflow dynamics 
where a suitable parameter associated with the spin flip probability was introduced \cite{godreche,krupa}.
The Ising Glauber and Sznajd models can be obtained by choosing specific values 
of the parameters in the generalised models with inflow and outflow dynamics respectively.

The exit probability plays an important role in the debate as it shows a marked difference in behaviour for the two models:
 for the Ising Glauber model, EP is linear; $E(x) = x$ while  for the Sznajd model 
\cite{slanina,redner,cas} 
\begin{equation}
E(x) = \frac{x^2}{x^2 + (1-x)^2},
\label{sznajd}
\end{equation} 
a distinctly nonlinear function of $x$.

 Another version of a generalised  model with inflow and outflow dynamics 
was proposed more recently  \cite{cas}
in which the range $r$ of the interaction  was varied. 
The Sznajd model with range $r$ (S(r) model hereafter) showed a range independent behaviour of the exit probability; EP is given by eq. (\ref{sznajd}) for all $r$.
For the generalised Ising Glauber model with $r$ neighbours (G(r) model henceforth),
numerical simulations were made which showed very good fitting to the 
 form given in eq. (\ref{sznajd}) for $r=2$ from which it was   claimed that non-linear behaviour 
of $E(x)$ can be observed  for inflow dynamics as well. 

\begin{figure}
\includegraphics[width=8cm,angle=0]{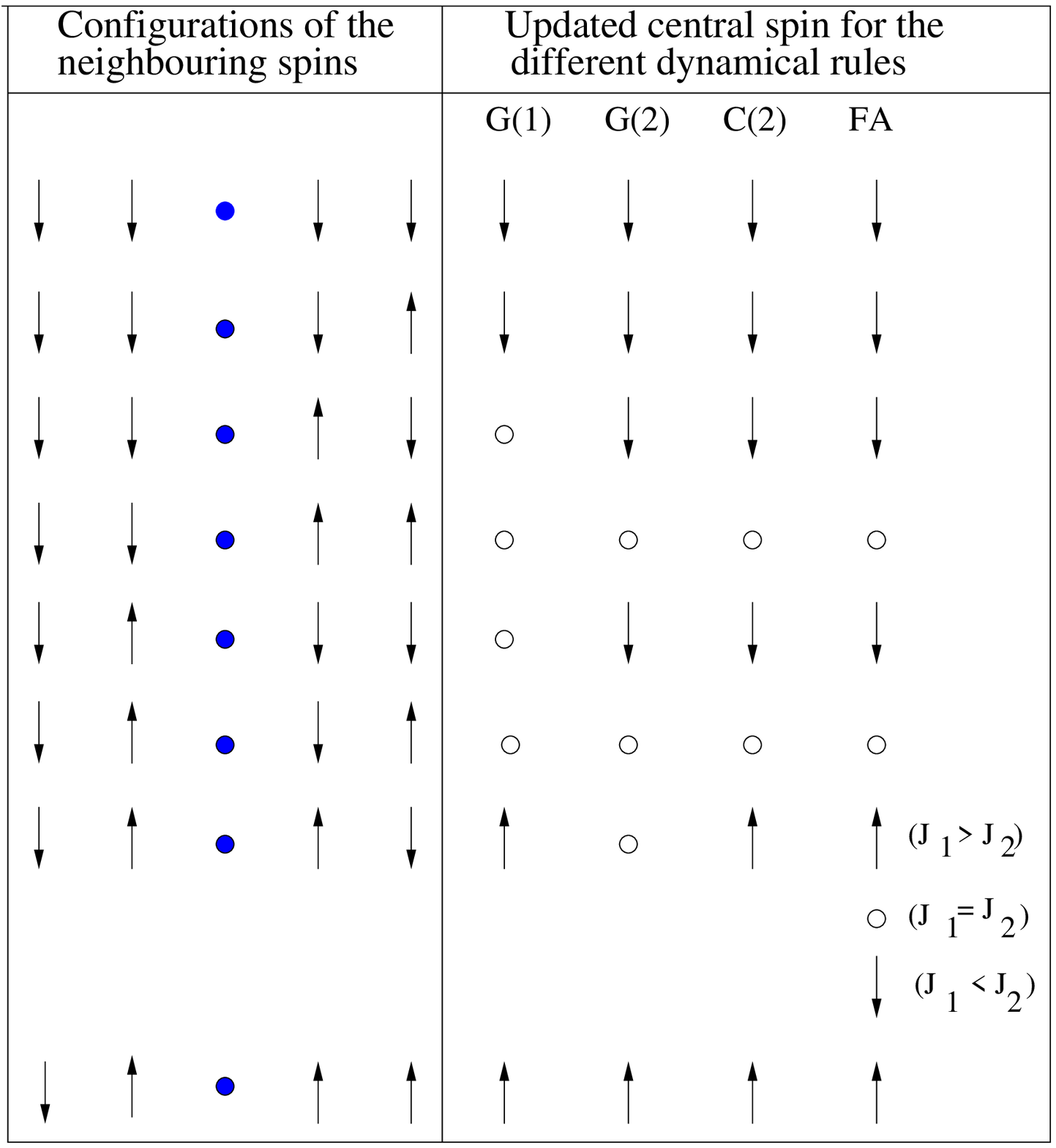}
\caption{(Color online) Left panel shows configuration of the neighbouring spins of the central spin represented by a  \textbullet
  which means either of the 
up/down states at time $t$. Right panel shows the state of the central spin after it is updated according to the 
different dynamical rules. A $\circ$ implies an ``undecided'' state  when
the up/down state occurs with equal probability.
The other eight states can be obtained by inversion.}
\label{table}
\end{figure}

A generalised $q$-voter model  which involves outflow dynamics 
has also been proposed \cite{qvoter}  in which $q$ neighbouring spins, 
if they agree, influence their other  neighbouring spins. 
In one dimension, $q =2$  corresponds to the Sznajd model and  the random version with $q=1$ (where only one of the two
boundary spins is updated  with equal probability)  
corresponds to the Ising Glauber/voter  model. 
The exit probability here again showed the property that it is independent of range. 

The shape of the exit probability is  an important issue.
Another interesting point to be noted is,  in all the different models studied so far \cite{cas,redner,slanina,qvoter} in one dimension, no finite size dependence has been 
noted in EP.  However,  there is a school of thought that such effects do exist 
and in reality exit probability has a step function behaviour 
in the thermodynamic limit \cite{galam} as observed in higher dimensions \cite{stauffer,networks,cas2d}. Such a step function behaviour also occurs for a 
special class of  one-dimensional
models 
where the dynamical rule involves the  size of the neighbouring domains \cite{bss,rbs}.
However, in the present work, we consider only those models with inflow dynamics (all of which are
short ranged) which belong to the Ising-Glauber class as far as dynamical behaviour
is concerned. 
Our aim is to find out how EP depends on various factors incorporated in the dynamics.
  Our main result is  
that a general form for the exit probability  given by  
\begin{equation}
E(x) = \frac{x^{\alpha}}{x^{\alpha} + (1-x)^{\alpha}}
\label{general}
\end{equation} 
exists, where   $\alpha$, the so called exit probability exponent 
 is very much dependent on factors like the  range of interaction, asymmetry 
of the model and fluctuation present in the dynamics.

\begin{figure}
\includegraphics[width=7.6cm,angle=0]{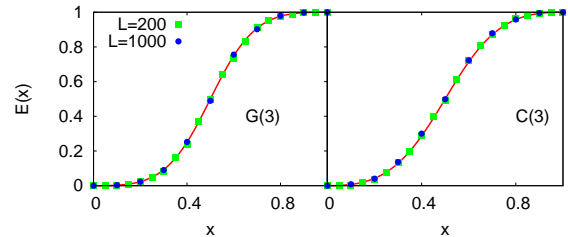}
\caption{(Color online) Exit probability versus initial concentration of up spins in the 
generalised Ising (G(r); left panel) and  cutoff models (C(r); right panel) with $r=3$. The $L=1000$ curves are  fitted with the best fit line of the form
given in eq. (\ref{general}).
}
\label{three}
\end{figure}

\noindent {\it Models and results}:  \\
1. The Ising Glauber model with $r$ neighbours (G(r)) :
 To update the $i$th spin $s_i(=\pm1)$ here, one  computes
\begin{equation}
 x = \sum_{j=1}^r [s_{i+j} + s_{i-j}].
\label{sum}
\end{equation}
If $x> 0, s_i = 1$,
if $x < 0, s_i = -1$ and $s_i$ is flipped with probability 1/2 if $x=0$.
For G(r), results are known for $r=1$ (exact) \cite{book} and $2$ (numerical) \cite{cas}. We have
 obtained results for higher values of $r$.\\
2. The cutoff model:  A model with a cutoff at $r$ called the C(r) model proposed in \cite{biswas-sen2} was also studied.  Here 
only the spins sitting at the domain boundary are liable to flip. To update such a  spin on site $i$, we calculate two quantities $r_1$ and $r_2$. 
$r_1$   
  is determined from the condition \\ 
$s_{i+1} =  s_{i+2}=\cdots= s_{i+r_{1}} \neq s_{i+r_1 + 1}$; \\
 and similarly $r_2$ is calculated from 
 the spins on the left side of the $i$th spin. $r_1$ and $r_2$ are both restricted to a maximum value $r$. 
Hence the neighbouring domain sizes  $r_1$ and $r_2$ 
are calculated subject to the  restriction that the  maximum size is $r$.
When $r_1$ is greater (less) than $r_2$, the state of the right (left) 
neighbours is adopted. If $r_1 = r_2$, the spin is flipped with probability 1/2.
C(r) is equivalent to G(r) for $r=1$.\\
3. The ferromagnetic asymmetric next nearest model  (FA) model:  
The G(2) or C(2) models can in fact be shown to be special cases 
of the Ising model with second neighbour interaction.
The Hamiltonian for this model is 
\begin{equation}
H = -J_1 \sum_i s_is_{i+1} -J_2 \sum_i s_is_{i+2}.
\end{equation}
 Here, the  role of asymmetry can be studied by varying $\kappa = J_2/J_1$. 
The special case $\kappa = 1$ is identical to G(2). $\kappa < 1$ corresponds to C(2) and for 
$\kappa > 1$ one may expect a different behaviour.
This system can be regarded as a ANNNI chain \cite{selke} with both interactions positive (ferromagnetic). By definition the FA model has range $r=2$.\\
4. The W(r)  model: 
 The W(r)  model is exactly like the Ising Glauber model except for the fact that when $x = 0$ in eq. (\ref{sum}), the spins are flipped with probability 
$W_0$ \cite{godreche}. It is known that for $W_0 =0$, which is called the constrained Glauber model, absorbing states are frozen states which are not the all up/down states. 
$W_0 = 0.5$ corresponds to the Ising Glauber model while $W_0 =1$ is the case of Metropolis rule.     
$W_0$  in a sense quantifies the fluctuation induced  by the dynamics, 
the fluctuation  is maximum when $W_0=1$ which causes the spins to flip
whenever $x$ in eq. (\ref{sum}) equals zero. 
In this model, we have studied the case for $r \geq 1$.  

In Fig. \ref{table}, we have presented the possible updated configurations 
for the central spin corresponding to eight  different configurations of its 4  nearest neighbours 
for G(1), G(2), C(2)  and FA.  The other eight cases can be obtained by inverting all the spins.
It is immediately noted  that G(r) and C(r) differ even for  $r=2$.
For FA, we may expect a new value of $\alpha$ for $\kappa > 1$, which,  however,  should not depend on the exact 
value of the $\kappa$.
It is also seen that the central spin is ``undecided'' in maximum number of cases in G(1), such cases are less in number for G(2) and even less in 
C(2) and  FA with $\kappa \neq 1$.  We will discuss the effect of this feature on 
EP later.

\begin{figure}
\includegraphics[width=6.5cm,angle=0]{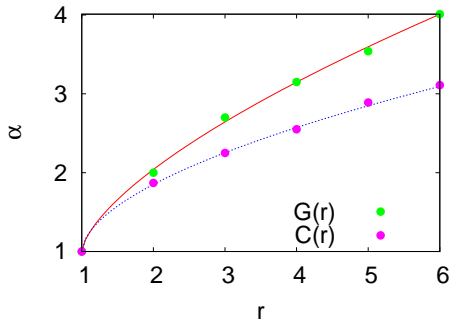}
\caption{(Color online) Plot of the exponent $\alpha$ against range $r$ for $G(r)$ and $C(r)$ models. The full line corresponds to the fitting form of eq.(\ref{equation})}.
\label{alpha}
\end{figure}

As mentioned before, the EP follows a behaviour given by eq. (\ref{general}) 
in all cases. Typical variation of the EP for G(r) and C(r) for $r=3$ 
are shown in Fig. \ref{three}. In Fig. \ref{alpha}, we plot the values of $\alpha$ against $r$ for these two models. We note that $\alpha$ is an increasing function of $r$ for both models. 
Hence $\alpha$ for G(r) is greater than 2 for $r > 2$ and the value of $\alpha = 2$ coincides with the S(r) value only for $r=2$. 
On the other hand, $\alpha$ for C(r) is less compared to G(r) for all $r > 1$. 
We try a general form to fit $\alpha$ with $r$  as

\begin{equation}
(\alpha - 1) = a(r-1)^b
\label{equation}
\end{equation}
and note that it shows a fairly good fitting for both G(r) and C(r) with 
$ a=1.04 \pm 0.02$, $b=0.66 \pm 0.02$ for G(r)
and $a=0.85 \pm 0.01$, $ b=0.56 \pm 0.01$ for C(r).
 Both $a$ and $b$ are larger for G(r) 
indicating the stronger dependence on $r$.   


\begin{figure}
\includegraphics[width=7.0cm,angle=0]{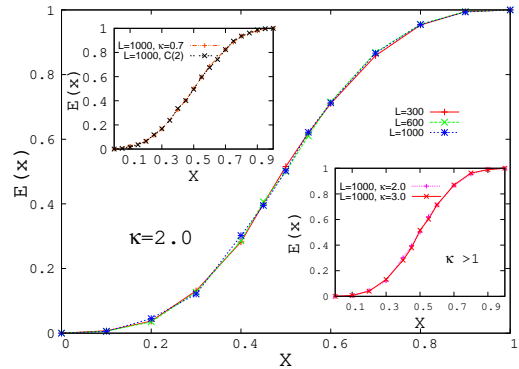}
\caption{(Color online) Exit probability for the FA model: main plot shows the result
for $\kappa =2$ for different system sizes. Top left inset shows $E(x)$ 
for $\kappa < 1$ and C(2) which give identical results as expected; bottom right inset 
shows that for $\kappa > 1$, EP is independent of the exact value of $\kappa$. The solid lines are guides to the eye.}
\label{FA}
\end{figure}
The FA model, as expected, gives $\alpha = 1.85 \pm 0.03$ for $\kappa  <  1$ which is identical to the C(2) value ($1.85 \pm 0.02$)   
and $\alpha = 2$ for $\kappa = 1$ (the G(2) model).
In the third case $\kappa  > 1$,  
we get a new value of $\alpha = 2.24 \pm 0.04$.
The results are shown in Fig. \ref{FA}.

\begin{figure}
\includegraphics[width=7.6cm,angle=0]{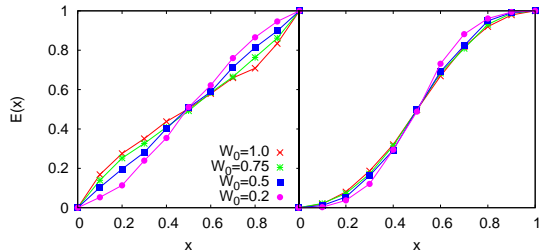}
\caption{(Color online) Exit probability versus initial concentration of up spins in the 
 W(r)  model, $r=1$ (left panel)  and $r=2$ (right panel). The solid lines are guides to the eye.}
\label{wr1r2}
\end{figure}

\begin{figure}
\includegraphics[width=5.5cm,angle=0]{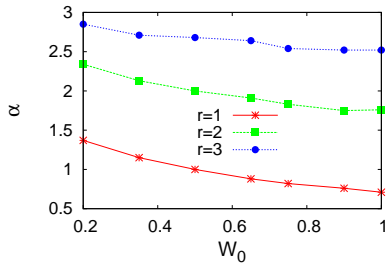}
\caption{(Color online) Exponent $\alpha$ against  $W_0$ in the W(r) model with 
$r=1$,  $2$ and $3$. The solid lines are guides to the eye.}
\label{wmodel}
\end{figure}

The W(r) model leads to  both qualitatively and quantitatively  different results. Even for $r=1$, the exit 
probability  does 
not have a linear dependence on $x$ for $W_0 \neq 0.5$; 
 $\alpha \neq 1$  unlike the Ising-Glauber case (Fig. \ref{wr1r2}).
Here too we find $\alpha$ to be dependent on $r$. 
We plot the dependence of $\alpha$ against $W_0$  for $r=1, 2$ and $3$ in Fig. \ref{wmodel}.
For the W(1)  model, $\alpha $ behaves as     $1/\sqrt{2W_0}$ as  $W_0 \to 1$. The values of $\alpha$ for $r=2$ and $r=1$ differ 
by unity for any $W_0$ as in the Ising Glauber model.  However, the 
differences in the values of $\alpha$ for W(3) and W(2)  weakly increase
with $W_0$. 
It is interesting to note here that Glauber ($W_0=0.5$) and Metropolis 
($W_0=1$) algorithms  give different values of $\alpha$ although for any $W_0 \neq 0$, the W(r)  model  belongs to
the Glauber universality class \cite{godreche}.

Some   general features can immediately be noted from the results. 
 If $r$ is increased, $\alpha$ increases indicating that the exit probability 
becomes steeper  in models 
with inflow dynamics.  
When $r$ is same in two models, $\alpha$ assumes different values due to the presence of other factors. For example, both $G(2)$ and FA ($\kappa > 1$) have $r=2$, but $\alpha$ is larger in the latter. 
The two models differ in the number of so called ``undecided states'' (see Fig. \ref{table}) and apparently $\alpha$ is larger when such states are less in number. 
In order to account for the fact that C(2) has a smaller value of $\alpha$ compared to G(2), although the 
number of undecided states is less here, one must also note that the effective number of neighbours in C(2) is less than 2. 
The combined effect makes the value of $\alpha$ smaller indicating that the range  has a stronger effect on EP than 
stochasticity. 

The results in the W(r) model can be qualitatively explained. 
For $r =1$, we note that the EP curves have different curvature for $W_0$ below and above $W_0= 0.5$. Let us take the case when $x < 0.5$ where EP 
is larger for $W_0 > 0.5$ compared to the value at $W_0 = 0.5$. This happens 
since the initial state here contains more spins in the down state, and the flipping probability 
is larger than 1/2. Same logic explains why EP is less when $x > 0.5$. 
At $x = 0.5$, $E(x)$ is equal to 1/2 for all models as $E(x) + E(1-x) =1$.
So the curves cross at $x = 0.5$ and 
 $\alpha$ has a smaller value than 1  for $W_0 > 0.5$ and larger value than 1 for $W_0 < 0.5$ (as for $W_0 = 0.5,~ E(x) = x$, or $\alpha =1$). 
$W_0$ effectively    controls  the fluctuation 
and we find that it can alter the value of $\alpha$. 
For larger values of $r$, similarly,  $\alpha$ is   larger (smaller)  than  the 
 G(r) values for $W_0 < 0.5$ ($W_0 > 0.5$). However, the curvatures 
are same as $\alpha> 1$ always.

We also note that  no system size dependence of the EP is observed  in any of the models 
even when $r$ is increased,  
asymmetry is introduced or fluctuation is modified. So no indication of a step function like 
EP is there for finite values of $r$ even in the thermodynamic limit. 
However, as $r$ is made larger, $\alpha$ increases and one can conclude that in the fully connected model
corresponding to the infinite dimensional case, $\alpha$ will diverge giving rise to a step function behaviour in the EP 
at $x = 1/2$.  

Some of the issues discussed in the beginning   may   be addressed now. First of all, 
it is evident that EP shows range dependence in models with inflow of information in general 
 in contrast to models with outflow of information, where 
increasing the range only results in change in timescales. 
In inflow dynamics, increasing $r$ apparently makes the system 
approach higher dimensional behaviour, although no system  size dependence 
appears. 
The fact that EP for S(r)  and G(2) model \cite{cas,qvoter,slanina} shows 
identical   behaviour ($\alpha =2$) 
seems to be  purely  accidental; there are inflow and outflow models with $r=2$ 
which have   $\alpha \neq 2$. However, $\alpha$ can be nonintegral
in inflow dynamics in contrast to known models with outflow dynamics \cite{cas,qvoter} ($\alpha = q$ for the $q$ voter model).  

An important issue is the question of universality. As already mentioned, 
all the models studied  here have the same dynamical behaviour as far as coarsening is concerned; they all belong to the Ising-Glauber class
with the dynamic exponent and persistence exponent identical. In fact, even the 
models with outflow dynamics like the Sznajd model belongs to this universality class \cite{stauffer1} (we have checked for $r=2$ as well). Thus we find that the exit probability is a nonuniversal quantity, it depends on the details of the dynamical rule and 
is not simply determined by the fact whether information flows out or in.
However, it seems safe to make the statement that  there is a clear difference: outflow dynamics
is characterised by no range dependence while inflow dynamics is. 

The question that may naturally arise after this discussion is why does the EP behave differently when the coarsening behaviour is identical. Here it should be remembered that coarsening behaviour is strictly relevant to a completely random initial configuration corresponding to $x = 1/2$. Indeed, at $x = 1/2$, 
in all the cases $E(x) = 1/2$. Hence a deviation from  the perfectly random 
state results in  
reaching  the all up/down states with different probabilities for the different models.  


%
%
%
%

In summary, we present evidence that the exit probability 
can be expressed in a general form. 
An exponent 
 $\alpha$ associated with the EP is identified which 
is  
 strongly dependent on the details of the system as far as inflow dynamics is concerned.
$\alpha$ can have nonintegral values (even less than unity) for inflow dynamics while for
the models with outflow dynamics studied so far, only integral values 
have been obtained. 
Most of  the observed results can be qualitatively explained. 
 
The range dependence distinguishes the inflow dynamics from outflow dynamics. Apart from the range dependence, the role of other factors in the dynamical rules also show their effect on 
EP in inflow dynamics.  
Effect of these factors in outflow dynamics may  bring out further 
distinguishing features, a study in progress \cite{rb}.

Acknowledgement: PR acknowledges financial support from  UGC. PS acknowledges financial support from CSIR project. S.B. thanks the Department
of Theoretical Physics, TIFR, for the use of its computational resources.

\end{document}